\documentclass[12pt,amsmath,amssymb,preprint]{revtex4}
\usepackage{graphicx}
\usepackage{bm}
\usepackage{setspace}
\begin{document}

\title{Spectra of secondary electrons generated in 
water by energetic ions}


\author{ Emanuele Scifoni$^1$\footnote{E-mail:
scifoni@fias.uni-frankfurt.de},
Eugene Surdutovich$^{1,2}$ and Andrey V. Solov'yov$^{1}$\footnote{E-mail:
solovyov@fias.uni-frankfurt.de}}

\affiliation{
$^1$Frankfurt Institute for Advanced Studies,
Ruth-Moufang-Str. 1, 60438 Frankfurt am Main, Germany\\
$^2$Department
of Physics, Oakland University, Rochester, Michigan 48309, USA}

\begin{abstract} 
The energy distributions of secondary electrons produced by energetic 
carbon ions (in the energy range used, e.g., in hadron therapy) , incident on liquid water, are discussed.
For low-energy ions,
a new parameterization of the singly-differential ionization cross
sections is introduced, based on tuning the position of the Bragg peak. 
The resulting parameterization allows a fast calculation of the energy
spectra  
of secondary electrons at different depths along the ion's trajectory, 
especially near the Bragg peak. 
At the same time, this parameterization provides penetration depths
for a broad range of
  initial-ion energies within the  therapeutically-accepted error.  
For high-energy ions, the energy 
distribution is obtained with a use of
the dielectric response function approach. 
Different models are compared and discussed. \end{abstract}


\maketitle

\section{Introduction}

An accurate energy distribution of electrons produced as a result of
the ionization of water molecules induced by fast incident ions is a key
piece of information for many applications. One of the most
important 
among these applications, since liquid water is considered to be a
good tissue-like medium, is related to its relevance to the physics of
ion-beam cancer therapy and radiation protection in space.  Ion beam
cancer therapy has become an operative treatment tool, whose power is
based on theoretical and computational methods of basic
science~\cite{Kraft05}.  On the other hand, radiation damage by ions
is becoming a major topic of research for shielding of human space
missions by the ESA and NASA~\cite{CucinottaDurante}.  These two fields of
applications differ slightly in the initial energies of incident ions, hundreds
of MeV/u for cancer therapy and $\sim$1~GeV/u for radiation protection
from galactic cosmic rays.

The goal of studies applied to both curing and shielding purposes is
an understanding and controlling of the radiation damage, which is
focused on DNA damage and its repair. Many details of the involved
physical processes, starting from the incidence of an energetic ion on
tissue and leading to biological damage, are still far from being
explained and calculated on a nanoscopic level.  An attempt to build
an inclusive picture of relevant physical processes was suggested in
our previous work \cite{us-ms, us-epn}, which proposed a multiscale
approach to this problem.

It is widely accepted that the major part of damage done by ions is
related, directly or indirectly, to the secondary electrons
(frequently called $\delta$-electrons) produced by ionization of the
medium. These electrons may interact with parts of DNA molecules
in the cell nuclei, generate other secondaries, such as other
electrons or radicals, which can then interact with DNA.  Besides this, electrons carrying energy, with their propagation define a volume around the track  of abrupt temperature increase that can lead to a local heating induced damage \cite{toule}; furthermore each ionization produces a hole which may also cause DNA damage.

In the multiscale approach~\cite{us-ms}, the energy spectrum of
secondary electrons occupies the key position. It is also an important
input for numerous Monte Carlo (MC) simulations of track
structure~\cite{Goodhead06,Nikjoo99,Nikjoo06}.  An increasing
interest for a more accurate shape of this distribution, especially at
low energies, has followed the well known experiments by the Sanche's
group~\cite{Sanche2000} revealing possible damage induced by
low-energy ($\sim$5~eV) electrons.

The studies of the energy distribution of secondary electrons have been
carried out by different groups (e.g.~\cite{Dingfelder2000, Nikjoo06}). In a
majority of works, energetic electrons or protons are the primary
projectiles~\cite{Emfie05, Dingfelder2000}; while  a few
studies have been extended to heavier ions~\cite{Ueh02a,Pimblott}.  In
our analysis, we consider carbon ions because they have been largely
used in ion-beam therapy both in Germany and Japan~\cite{Hiroshiko}.  However, our
analysis can also be extended to other ions.

The difficulties in analyzing singly differential ionization cross
sections are basically of two types: (1) the impossibility to depict all
energy ranges using the same formalism because the energy range of
interest is too large to be included in the range of validity of a single
approximation; (2) the deviation of the ion's propagation through 
liquid water from that of water vapor
often used in experiments, is unknown.  As the
applicability of the Born approximation is a key issue in this
connection, we divide our analysis (as it is usual in similar
works~\cite{Dingfelder2000, Nikjoo06}) between the fast and relatively
slow regimes of the projectile's velocity, corresponding to different
  parts of the ion's trajectory.
First, we start with a general approach valid for slow ions. This
approach is the only possibility in that energy range, and the 
accuracy in the shape of
secondary electron spectra is limited in the entire energy range.  
Then, we present a method valid only for fast ions, but
  returning more accurate electron energy distributions in that range.

\section{Slow regime} 
In order to account for the region  where the ion's energy is smaller
  than 500 keV/u, where the Born approximation 
is not valid, we use a parametric approach built on existing
experimental data.  As was shown in our previous papers
\cite{nimb,epjd}, the probability to produce $dN$ secondary electrons
with kinetic energies from $W$ to $W+dW$, $\frac{dN}{dW}dW$, from a
segment of the ion track $d\zeta$, is proportional to
$\frac{d\sigma}{dW}d\zeta$, where $\frac{d\sigma}{dW}$ is the singly
differential ionization cross section (SDCS). This cross section is
the main characteristic in our analysis defining both the ion's
stopping in the medium and the energy spectrum of secondary electrons.

Since the angular distribution of prevailing low-energy electrons
($W<45$~eV) is rather flat over the whole angular range~\cite{Tob80},
the SDCS is sufficient for our analysis.  Besides the
kinetic energy of secondary electrons and the properties of water
molecules, the SDCS depends on the kinetic energy of projectiles, $T$,
and their charges, $z$.

\subsection{Semi-empirical Model for vapor water}
\label{ionpass} 

In our previous papers~\cite{nimb,epjd}, we used a model based on a parameterization by Rudd~\cite{Rudd92}, which yields an analytical formula for SDCS on a wide energy range combining the experimental data for water vapor,
calculations within the plane-wave Born approximation, and other
theoretical models.  This model, original for protons, has been
extended first to helium~\cite{Ueh02a} and then to heavier
ions by us~\cite{nimb,epjd} and a few other works~\cite{Plante}.

We used the following parametric functions~\cite{epjd}:
\begin{eqnarray} \frac{d\sigma (W,T)}{dW}=
  z_{eff}^2 \sum\limits_i \frac{4\pi a_0^2 N_i}{I_i}
  \left(\frac{R}{I_i}\right)^2 \times \\ \nonumber
  \frac{F_1^i(v_i)+F_2^i(v_i)\omega_i}{\left(1+\omega_i\right)^3\left(1+\exp(\alpha^i
    (\omega_i-\omega^{\rm max}_i)/v_i)\right)}~,\\ \label{sdcs} F_1^i(v)
  = A_1^i \frac{\ln(\frac{1+v^2}{1-\beta^2})-\beta^2}{B_1^i/v^2+v^2} +
  \frac{C_1^i v^{D_1^i}}{1+E_1^i v^{D_1^i+4}}~,\\ \label{f1} F_2^i(v) = C_2^i
  v^{D_2^i} \frac{A_2^i v^2+B_2^i}{C_2^i v^{D_2^i+4}+A_2^i v^2+B_2^i}~. \label{f2}
  \end{eqnarray} 
\noindent where the sum is taken over the electron shells of the water
molecule, $a_0$ is the Bohr radius, $R$ is the Rydberg, $N_i$ the
shell occupancy, $I_i$ the ionization potential of the shell,
$\omega_i=W/I_i$~, $v_i=(m V^2/(2 I_i))^{1/2}$, $\omega^{\rm max}_i =
4v_i^2-2v_i-R/(4I_i)$; $m$ is the mass of electron, $V$ is the
velocity of the projectile, $T$ its kinetic energy and $\beta=V/c$.

The corresponding fitting parameters, taken from Ref.~\cite{Rudd92},
$A_1^i$ ... $E_1^i$, $A_2^i$ ... $D_2^i$, $\alpha^i$ are listed in
Table~\ref{param} and the ionization potentials in Table~\ref{pot}
(vapor row).
\begin{table}
\caption{Fitting parameters for the two inner shells of the water
  molecule ($1a_1$, $2a_1$)  and 
  for the three outer shells ($1b_2$, $3a_1$, $1b_1$) in the original parameterization by
  Rudd for vapor~\cite{Rudd92} and in a new parameterization here introduced for liquid.}
        {\begin{tabular}{cc|cccccccccc} \hline\noalign{\smallskip}
            & parameter & $A_1$ & $B_1$ & $C_1$ & $D_1$ & $E_1$ & $A_2$
            & $B_2$ & $C_2$ & $D_2$ & $\alpha$ \\ \hline inner shell & &
            1.25 & 0.5 & 1.0 & 1.0 & 3.0 & 1.1 & 1.3 & 1.0 & 0.0 &
            0.66 \\ \hline outer shell & vapor~\cite{Rudd92} &0.97 & 82 & 0.4 & -0.3 & 0.38 & 1.04
            & 17.3 & 0.76 & 0.04 & 0.64 \\ \hline outer shell & liquid (this work) &1.02 & 82 &
            0.50 & -0.78 & 0.38 & 1.07 & 14.5 & 0.61 & 0.04 &
            0.64\\ \hline
\end{tabular}} \label{param} \end{table} 
\begin{table} \caption{
Ionization potentials for vapor~\cite{Rudd92} and liquid~\cite{Dingfelder} water (in eV). In the two approaches presented in this paper, the first and the second set are respectively used.}
  {\begin{tabular}{c|c|c|c|c|c} \hline shell & $1b_1$ & $3a_1$ &
      $1b_2$ & $2a_1$ & $1a_1$ \\ \hline vapor &12.61 & 14.73 & 18.55
      & 32.2 & 539.7 \\ \hline liquid &10.79 & 13.39 & 16.05 & 32.3 &
      539.0 \\ \hline
\end{tabular}} \label{pot} \end{table}

This parameterization is different from the original~\cite{Rudd92} in
two respects: first, the expression for $F_1$ is modified so that it
has the correct asymptotic behavior in the relativistic
limit~\cite{epjd}, and second, it contains an effective charge of the
ion $z_{eff}$, which depends on the velocity of the ion. The effective
charge takes into account the effect of charge transfer, i.e., gradual
reduction of the original charge of the ion as it slows down. For the
effective charge, we have used an expression given by
Barkas~\cite{Barkas63}, $z_{eff}=z(1-\exp(-125\beta z^{-2/3}))$.

These two modifications give the shape of the LET dependence along the
track for a single ion. However, ions in the beam are spread in
energies due to scattering. The straggling in energies is taken into
account in Ref.~\cite{epjd}, but more details are given in
Ref.~\cite{RADAM2}.  All in all, our resulting general parameterization of SDCS for water yielded a 97\% accuracy in reproducing the
Bragg's peak position in comparison with experiment and
compared very well with MC simulations~\cite{Pshen} in shape (without
nuclear fragmentation).

\subsection{Semi-empirical model extensions for liquid water} 

However, the parameterization for SDCS, used in
Refs.~\cite{epjd,RADAM2}, did not take into account medium density
effects, i.e., the effects arising from the difference between the
ion's propagation in liquid water from that of vapor. We
used the liquid-water density, but took all other parameters
corresponding to water vapor. Unfortunately, the experimental data for
ionization cross sections in
liquid water are very limited and insufficient for making an
independent parameterization. However, since the SDCS of
  ionization along with excitation contribute to the stopping cross
  section, directly related to LET~\cite{epjd}, it is
  possible to extract a modified parameterization of the SDCS for
  liquid water from different available experiments. In
  Ref.~\cite{Dingfelder2000}, the experimental measurements of
  stopping cross sections~\cite{ICRU93} are used as a function of the
  ion's 
  energy. This only provides a single curve to which to compare.  
Instead, we choose to fit the parameterization
with measurements of the LET dependence on the
  penetration depth in liquid water.  These depths have been measured at
  GSI~\cite{Schardt4, Schardt}. This not only provides several curves at
  different initial energies, but also a direct comparison with the
  carbon ion data.  
So instead of tuning parameters directly
  to the stopping cross sections as in \cite{Dingfelder2000}, we
  use the reported Bragg peak profiles, for different initial ion energies
  and thus the LET as a function of the penetration depth.

 In fact, the integration of the inverse of the LET gives the position of
the Bragg peak. Hence, we can adjust the parameters of the SDCS by tuning the
depth of the Bragg peak to the experiment.  Such a procedure is rather
simple and fast using our approach~\cite{epjd}.

So, we start with the SDCS of ionization, which includes relativistic
effects and charge transfer as an input to the stopping cross section,
defined as
\begin{eqnarray} \sigma_{\rm
st}=\sum\limits_i \int^{W_{max}}_0 (W+I_i)\frac{d\sigma_i(W,T)}{dW}dW~.
\label{eq6} \end{eqnarray} 
Now, we need to add the contribution of excitation to the stopping
cross section. These cross sections are accounted by scaling with our effective charge, depending on ions' velocities, a
semi-empirical formulation by Miller and Green \cite{MillGre} for protons
\begin{eqnarray} \sigma_{exc, k}(T)~=z_{eff}^2\frac{a_k(T-E_k)}{b_k+T^{c_k}},
\label{eq6ast} \end{eqnarray} 
where the index $k$ corresponds to five different excitation
transitions, i.e., 
to states A$^1$B$_1$, B$^1$A$_1$, Rydberg A+B, Rydberg
C+D, and diffuse bands, respectively;  
$\sigma_{exc, k}(T)$ is the excitation cross
section, $E_{exc, k}$ is the corresponding excitation energy and
$a_k,b_k,c_k$ are parameters reported in Table III and computed from data reported and fully explained in Refs.\cite{Dingfelder2000, MillGre}, giving cross sections in $\AA^2$ when energy is input in eV. The stopping cross section is then calculated~\cite{epjd}
as
\begin{eqnarray} 
\sigma_{\rm st}^\ast(T)=\sigma_{\rm st}(T)+\sum\limits_iE_{exc,
  i}\sigma_{exc, i}(T)~. \label{eq6ast}
\end{eqnarray} 
The LET is given by $n\sigma_{\rm st}^\ast(T)$, where $n$ is a number
density of liquid water molecules.
\begin{table} \caption{Energies and parameters for excitation cross sections (Eq.~\ref{eq6ast}) corresponding respectively to states  A$^1$B$_1$, B$^1$A$_1$, Rydberg A+B, Rydberg C+D, and diffuse bands.}
  {\begin{tabular}{c|c c c c} \hline
  $k$ & $E_k(eV)$ & $a_k$& $b_k$ & $c_k$ \\ \hline
1&  8.17& 2245& 4493 & 0.85 \\
2& 10.13& 6319& 7020 & 0.88\\
3& 11.31& 4387& 8135  &0.88\\
4& 12.91& 989& 3172 & 0.78\\ 
5& 14.50& 1214& 3352 & 0.78\\
 \hline
 
\end{tabular}} \label{exc} \end{table}

\begin{figure}
\includegraphics[width=.95\textwidth]{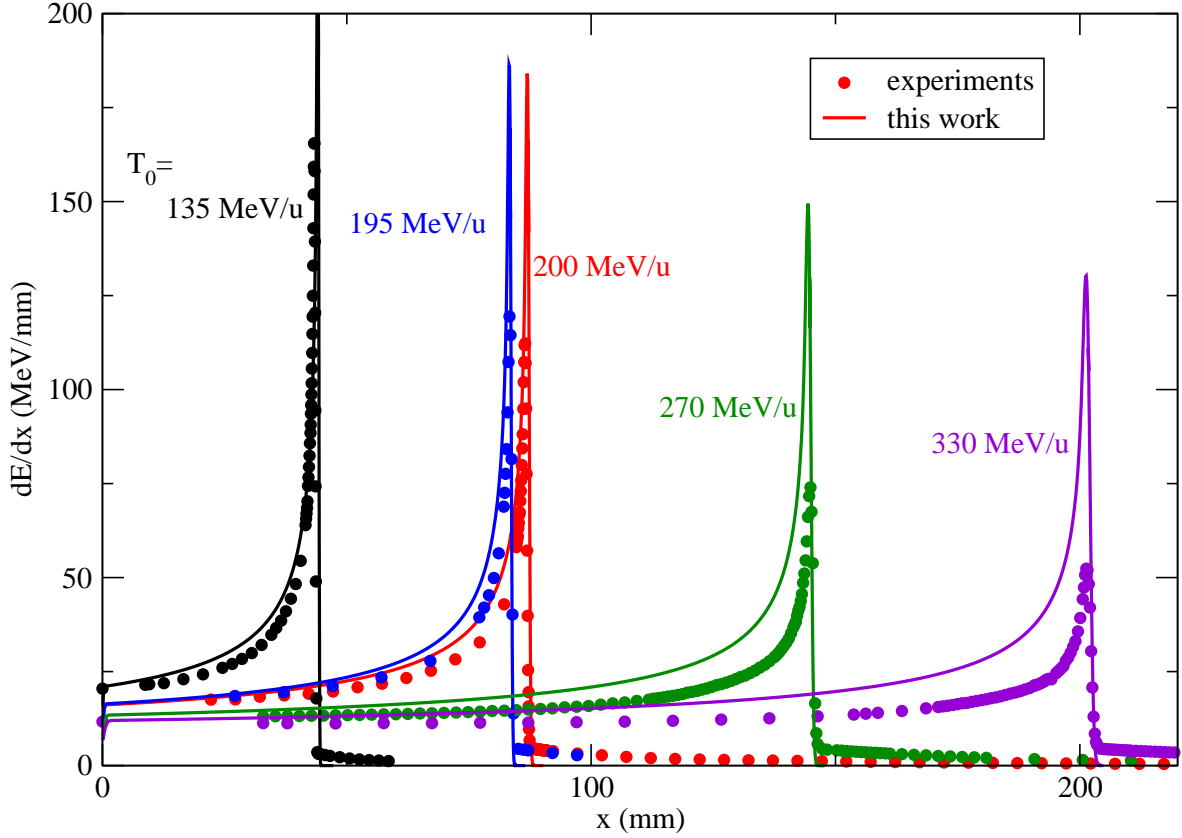} \caption{Linear energy
  deposition for carbon ions for different initial energies T$_0$, with
  our model compared to experiments from GSI~\cite{Schardt4, Schardt}.}
\label{4peaks} 
\end{figure}

At this point, we add the energy straggling correction, i.e., a
widening of 
the peak arising from the stocasticity of the energy loss phenomenon and related to the multiple scattering of ions. This correction
affects also the position of the Bragg peak, slightly reducing its depth. The
correction is given by
\begin{eqnarray} 
\lefteqn{\left\langle \frac{dT}{dx}(x)\right\rangle =}\\ \nonumber
&&\frac{1}{\sigma_{str}(x_0)\sqrt{2\pi}}\int_{0}^{x_0}
\frac{dT}{dx}(x')\exp
(-\frac{(x'-x)^{2}}{2\sigma_{str}(x_0)^{2}})dx'~, \label{eqstrg}
\end{eqnarray} 
where $x_0$ is a maximum penetration depth of the projectile, and
$\sigma_{str}(x_0)=0.012x_0^{0.951}/\sqrt{A}$ is the
longitudinal-straggling standard deviation (in cm, when $x_0$ is also in cm) computed with a
phenomenological formula introduced by Chu \cite{Chu}, and depending
(through $x_0$) on the initial energy of the ion $T_0$. $A$ is the
ion's mass number.
\begin{figure*}
  \includegraphics[width=.95\textwidth]{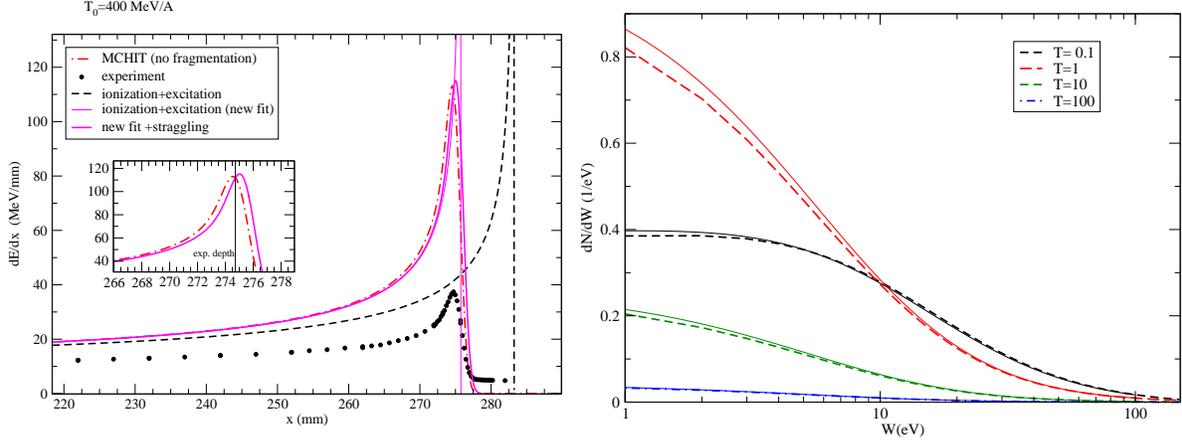} \caption{(Color
    online) Linear
    energy deposition (left panel) for a carbon ion with incident
    energy of 400 MeV/u, computed with different parameterizations,
    within the experimental error of the original data, and compared to
    experiments~\cite{Schardt} and MC simulations~\cite{Pshen} (where no nuclear fragmentation is considered). The zoom in the inset shows also the experimental penetration depth. On the right panel the effect of this new
    parameterization (solid line) on the spectra of secondary electrons
    compared to the previous one (dashed line).}
\label{tune}
\end{figure*}
We found, that if we gradually change the parameters from Rudd's
model within the experimental error~\cite{Rudd92}, the Bragg peak
position can be tuned to match those obtained in stopping power
experiments (see Fig.~\ref{tune}), as well as MC transport code simulation (MCHIT code based
on GEANT4 toolkit~\cite{Pshen,PshPC}). The tuned parameters are shown in the
third row of Table~\ref{param}.  The same parameterization is
applicable to a broad range of energies (see Table~\ref{tab3} and
Fig.~\ref{4peaks}), keeping the Bragg peak
position within the acceptable error for therapeutic uses (0.5 mm).
\begin{table} 
\caption{Bragg peak positions for different incident energies of
  carbon ions in water obtained with the present parameterization with
  and without accounting for straggling and compared with
  experiments~\cite{Schardt4, Schardt} and Monte Carlo simulations~\cite{Pshen,PshPC}.}
\begin{tabular}{|c||c|c|c|c|} \hline T(MeV/u) &
\multicolumn{4}{c|}{Bragg peak position (mm)}\tabularnewline \hline &
us (single ion) & us (straggling) & Schardt & MCHIT\tabularnewline
\hline \hline 400 & 275.75 & 275.01 & 274.72 & 274.53\tabularnewline
\hline 330 & 201.81 & 201.30 & 201.42 & 200.52\tabularnewline \hline 270 &
144.78 & 144.42 & 144.82 & 143.78 \tabularnewline \hline 200 & 87.19 & 86.94
& 86.46 & 86.90\tabularnewline \hline 195 & 83.49 & 83.26 & 83.39 &
\tabularnewline \hline 135 & 44.20 & 44.05 & 43.34 & 43.73\tabularnewline
\hline
\end{tabular} 
\label{tab3}
\end{table}
In particular, the comparison with Monte Carlo simulations,
 shows on one hand an almost perfect agreement in the shape of the curves when considering only the electromagnetic processes and no nuclear fragmentation channels; as shown in the limit case of a 400 MeV/u ion beam (left panel of Fig.~\ref{tune}), where fragmentation has a drastic effect on the peak shape, but nevertheless doesn't affect the depth. On the other hand it reveals a slightly better reproducibility of the experimental depths in the whole range of energies
for the present method (Table~\ref{tab3}), as compared to MC, which is  based on a single value of average excitation energy ($<I>$=85 eV).

Thus, the approach suggested in Ref.~\cite{epjd} with the new parameters given
in Table~\ref{param}, can be used successfully also for rapid calculations of the Bragg peak position; and, in case of limited nuclear fragmentations, as occur in smaller ions, may also provide a correct deposition profile depiction~\cite{EmaISACC09}

\section{Fast regime} 
When ions are more energetic than 500 keV/u, the Born approximation is
valid, and a direct connection with the experimental data
for liquid water is possible through the dielectric response model.
This provides an opportunity for a more accurate calculation of the SDCS
and thus the energy spectrum of secondary electrons.

\subsection{The dielectric response
model: optical approximation} In this model, presented by a number of
studies for electrons and protons~\cite{Dingfelder2000, Emfie03}, but rarely
extended to heavier ions~\cite{Pimblott}, the only parametric quantity is retrieved
from measurements of photoionization cross sections.  The main
advantage of this model, which makes it suitable, especially for
treating condensed media, is the simultaneous accounting for both
single-particle and collective effects in the analysis of the response
of a medium to the ionizing/exciting energetic particle.  The key
quantity is the energy loss function, often denoted as $\eta_2$,
derived by the dielectric function $\epsilon$ and directly connected
to the generalized oscillator strength $df/dE(E,k)$\cite{landau8, Inokuti}:
\begin{equation} \eta_2(E,k) ={\rm
Im}\left(\frac{-1}{\epsilon(E,k)}\right)=
  \frac{\pi\Omega_p^2}{2ZE}\frac{df(E,k)}{dE}~,
\label{elf} \end{equation} 
where $Z$ is the total number of electrons of a molecule of the
medium, $E$ is the transfered energy, $k$ is the transfered momentum,
 and $\Omega_p=4R(\pi a_0^3n Z)^{1/2} =21.46$~eV is the plasma
  frequency computed for liquid water where $Z=10$.

Then, for a liquid medium, it is more convenient to insert a
macroscopic cross section $\Lambda=n\sigma$, and consider \cite{Inokuti}
\begin{equation} 
\frac{d\Lambda}{dEdk}=\frac{z_{eff}^2m_p}{\pi
  a_0T}\frac{\eta_2(E,k)}{k}, \label{integral}
\end{equation} 
where the factor $m_p$ is the mass ratio between proton and electron,
and, after integration over $k$
\begin{equation}
\frac{d\Lambda}{dE}=\frac{z_{eff}^2m_p}{\pi
  a_0T}\int_{k_{min}}^{k_{max}}\eta_2(E,k)\frac{dk}{k}~. \label{integral1}
\end{equation}
This can be expanded under the condition $E<<T/m_p$ in the following sum
\begin{equation} \frac{d\Lambda}{dE}\simeq\frac{z_{eff}^2m_p}{\pi
a_0T}\left[\frac{\eta_2(E,0)}{2}ln\frac{T}{m_pR}+ B(E)+O(E m_p/T)\right]~.
\label{botetc} 
\end{equation}

\begin{figure} 
\includegraphics[height=.4\textheight]{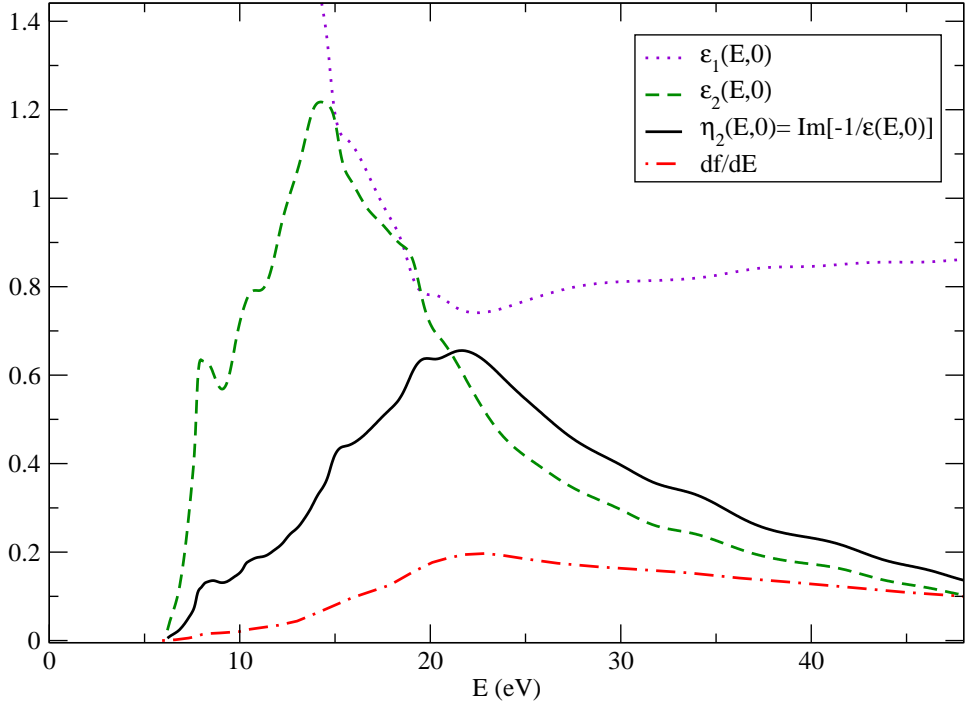}
\caption{Dielectric function data retrievable by optical experimental
  data~\cite{Hayashi00}.} \label{osc}
\end{figure}

In general, $\epsilon(E,k)=\epsilon_1(E,k)+i\epsilon_2(E,k)$, where
both the real part $\epsilon_1$ and the imaginary part $\epsilon_2$
are derivable from experiments~\cite{Hayashi00}. They are shown in
Fig.~\ref{osc} 
in 
the optical limit ($k$=0). Based on these functions, it is possible to
derive $\eta_2$ vital for our calculations.

 Thus, in cases where the optical limit is reasonable, i.e., when it
 is possible to consider ionization by ions to be similar to that of
 photons, it is possible to get a good approximation of the
 macroscopic cross section by retaining only the first term of
 Eq.~(\ref{botetc}), which is called Bethe optical term~\cite{Emfie05}.

In passing from the energy loss to the secondary electron distribution,
it is necessary to introduce the different ionization thresholds $I_j$ (here the real liquid water data are used, second row of Table~\ref{pot}) 
and contributions from different shells
\begin{eqnarray}
\frac{d\Lambda}{dW}&=&\sum_j\frac{d\Lambda_j}{dW}\nonumber \\ \frac{d\Lambda_j}{dW}&\simeq&\frac{z_{eff}^2m_p}{\pi
  a_0T}\left[\frac{G_j(E)\eta_2(W+I_j,0)}{2}\ln\frac{T}{m_pR}+
  B(E)+O(E/T)\right]~.
\end{eqnarray} 
The factors $G_j(E)$ account for different contributions to the optical spectrum from ionization of different shells. They are computed as suggested in a series of studies performed for protons Ref.~\cite{Emfie03} by  treating separately the different transitions by decomposing the resulting optical data in a sum of Drude functions, with the constrainth of respecting the sum rule.
\begin{eqnarray}
\epsilon_2(E,0)=\sum^{ion}_j\epsilon^{(j)}_{2,ion}(E,0)+\sum^{exc}_i\epsilon_{2,exc}^{(i)}(E,0)
\end{eqnarray}
where the $\epsilon^{(j)}_{2,ion}$ and the $\epsilon_{2,exc}^{(i)}$ represent, respectively, the contributions of a given ionization or excitation mode to the total imaginary part of the dielectric function and are computed differently. In particular the ionization contributions are expressed well by conventional Drude functions:
\begin{eqnarray}
\epsilon^{(j)}_{2,ion}(E,0)=\Omega_p^2\frac{f_j\gamma_jE}{(E_j^2-E^2)+(\gamma_jE)^2}
\end{eqnarray}

where the parameters $f_j,\gamma_j,E_j$ are respectively oscillator strength, width and position of the given transition and are found by imposing the sum rules:
\begin{eqnarray}
\int_0^\infty E\epsilon_{2}(E,0)dE=\Omega_p^2\pi/2 
\end{eqnarray}
\begin{eqnarray}
\int_0^\infty E\eta_{2}(E,0)dE=\Omega_p^2\pi/2
\end{eqnarray}

\begin{figure}
\includegraphics[height=.4\textheight]{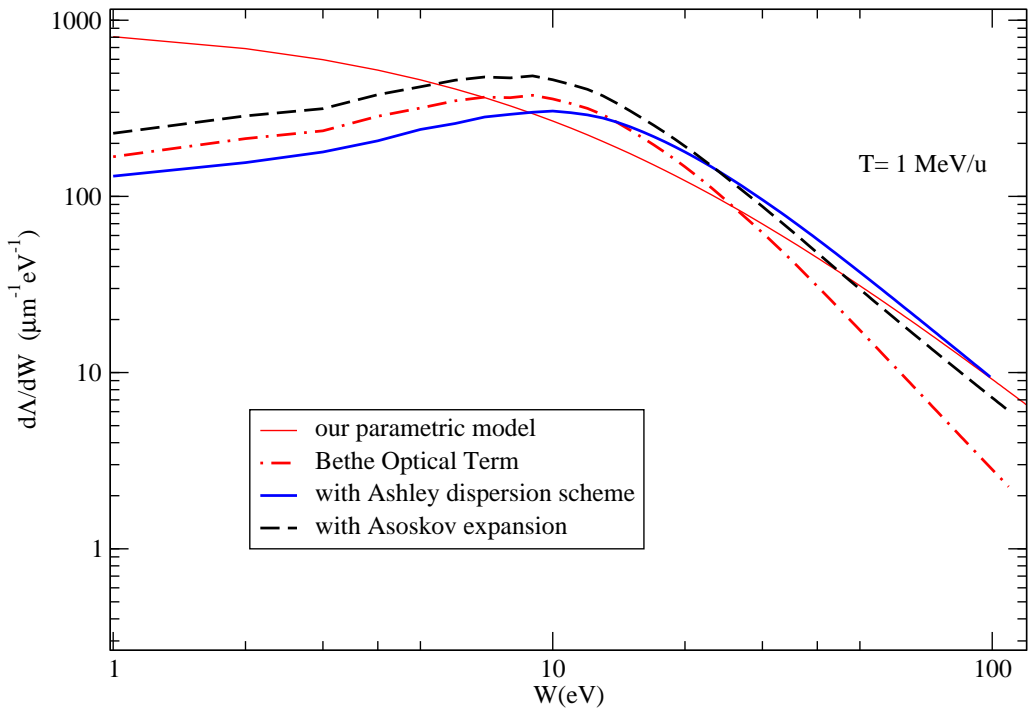}
\caption{Energy spectra of secondary electrons induced by a carbon
  ion: Comparison 
  between the parametric model and different dielectric response model
  approaches for liquid water.} \label{ashley}
\end{figure}

 We  use here the last fitted  parameters from Ref.~\cite{Emfie05} (see table 5), for the single contributions, while we directly derived $\epsilon_1$ and $\epsilon_2$ from experiments in the region 1-50 eV, and extrapolating with a non-linear curve fitting to the asymptotes (respectively 1 and 0).
\begin{table} \caption{Parameters (positions, widths and strengths) for the ionization transitions used for the shell contributions to $\epsilon_2$ \cite{Emfie05}}
  {\begin{tabular}{c|c c c} \hline

shell& $E_j$(eV) & $\gamma_j$ (eV) & $f_j$\\\hline
$1b_1$ & 16.30&  14.00&  0.2300\\
$3a_1$&  17.25&  10.91&  0.1600\\
$1b_2$&  28.00&  27.38&  0.1890\\
$2a_1$&  42.00&  28.68&  0.2095\\
$1a_1$& 450&  360&  0.3143\\\hline
\end{tabular}} \label{emfie}
\end{table}
Thus we take $G_j(E)=\epsilon^{(j)}_{2,ion}(E,0)/\epsilon_2(E,0)$.
  We can probe this approximation,
as mentioned, under the condition of $E\ll T/m_p$; this turns out to be
valid for relatively high $T$ ($>$0.5~MeV), when considering the lower
energy part of the secondary electron spectrum.  From
Fig.~\ref{ashley}, 
\begin{figure}
\includegraphics[height=.4\textheight]{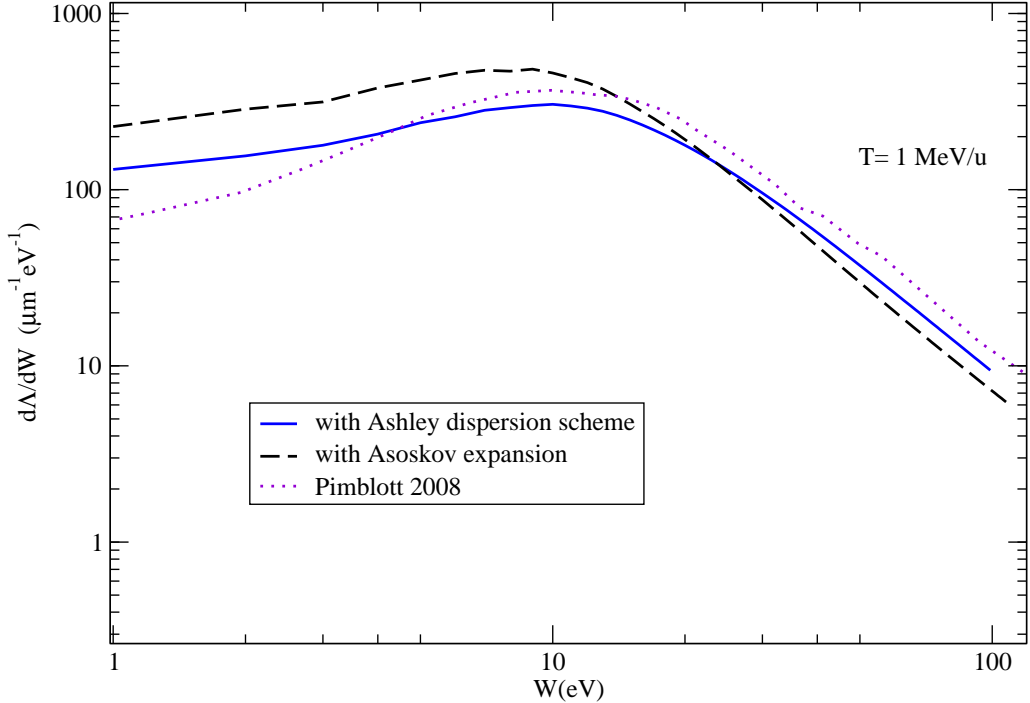}
\caption{Energy spectra of secondary electrons induced by a carbon
  ion: Comparison between the reported models and Ref.~\cite{Pimblott}.} \label{pimb}
\end{figure}
we can see that compared to our parametric formalism, there is a region
where the two approaches give very close results (10-40 eV), while at
a lower energy the curve corresponding to the parameterization deviates, as it is not able to correctly reproduce the liquid medium,
and at a higher electron energy
the Bethe optical approximation is not valid; thus, the
correct behavior is there described by the parametric, vapor-like curve.

\subsection{The dielectric response model: extension in the
energy-momentum plane} Finer approaches extend to $k>$0, i.e.,
generating the full energy-momentum plane, characterized by the
well-known Bethe ridge~\cite{Dingfelder}, through different models,
usually called dispersion schemes. These models follow the requirements to generate a generalized energy loss function $\eta_2(E,k)$ obviously  consistent in the optical limit ($k\rightarrow 0$) to the corresponding optical function, as well as in the asymptotic behaviour for large momentum transfer to the free electron limit:
\begin{equation}
\lim_{k\to\infty}\eta_2(E,k)\sim\delta(E-Q)~,
\end{equation}
where  $Q=k^2/2m$ is the free electron recoil energy \cite{Fern}. 
   Many methods have been proposed,
such as the one by Ashley~\cite{ashley}, based on a $\delta$-function representation of the optical oscillator strength of an atom

\begin{equation}
\frac{df(E)}{dE}=\sum_jf_j\delta(E-E_j)
\end{equation}
(with $f_j$ and $E_j$ respectively height and position of a given energy transition),  where $E_j$ is extended for $k>0$ to $E_j+Q$ and a generalization to condensed matter is performed:
\begin{eqnarray} 
\eta_2(E,k)=\int_0^\infty\frac{E'}{E}\eta_2(E',0)\delta(E-(E'+Q))dE'
=\frac{E-Q}{E}\eta_2(E-Q,0)\Theta(E-Q)
\end{eqnarray}
where $\Theta(E-Q)$ is a step function, equal to 1 for $E>Q$ and 0
otherwise. The $k$  dependence from the optical data is then achieved. 

Assuming that the weights computed  in the case of 0 momentum transfer don't change significantly, it is then possible to assume $G(E,k)=G(E)$
and compute the contributions to ionization of the shells with the eq.19
and
\begin{equation}
\frac{d\Lambda_j(E,k)}{dW}=\frac{z_{eff}^2m_p}{\pi
  a_0T}\int_{k_{min}}^{k_{max}}G_j(E)\eta_2(E,k)\frac{dk}{k}~. \label{integral11}
\end{equation}

 The integration in $k$, as shown in Eq.(\ref{integral1}),
is then performed with the minimum and maximum momentum transfer as
the integration limits:
\begin{eqnarray} 
\label{eq:ashley2}
k_\pm=\sqrt{2m}(\sqrt{T}\pm\sqrt{T-E}) \end{eqnarray}

Thorough reviews of the various possibility to expand in the Bethe surface are available in Refs.~\cite{Emfie05b, Dingfelder08}.
Other approaches include also a way for accounting the relativistic regime of the projectile energy. In general, this correction is small in the energy ranges where is the main focus of interest in evaluation of electrons spectra (Bragg peak region, i.e. $T<10MeV/u$), where $\beta^2<10^{-2}$. We report here, anyway, the results obtained by using such an approach, suggested by Asoskov et
al.~\cite{asos}, extended by us from an atomic single shell system  to include a multishell target. This formalism uses a different method  but perform still a similar expansion in the energy-momentum plane by
accounting for the recoil energy in order to express the generalized
oscillator strength (GOS) as a function of the optical oscillator
strength (OOS).  That, after integration in $k$, bring to the formula: \begin{eqnarray}
\label{eq:asoskov} \frac{d\Lambda}{dW}&=&\sum_j\frac{d\Lambda_j}{dW}~,\\
\frac{d\Lambda_j}{dW}&=&\frac{8 a_0^2R^2N_i z_{eff}
  Z}{\Omega_p^2mV^2}\left(\frac{E\epsilon^{(j)}_{2,ion}(E)}
     {E\epsilon^2}\left(\ln\frac{2mV^2}{E|1-
       \beta^2\epsilon|}-\frac{\epsilon_1-\beta^2
       \epsilon^2}{\epsilon_2}\arctan\frac{-\beta^2\epsilon_2}{1-\beta^2\epsilon_1}
   \right)+\frac{F(E)}{E^2}\right)\nonumber
\end{eqnarray}
where \begin{eqnarray} \label{eq:asoskov2}
  F(E)=\int_0^E\frac{E'\epsilon^{(j)}_{2,ion}(E')}{\epsilon^2}dE' \end{eqnarray}
and, for each shell, $E=W+I_i$.

In Fig.~\ref{ashley}, one can appreciate corrections induced by
these approaches to the Bethe optical term, which is reliable only for
very small values of $W$, as expected. We can see how the parametric and other
curves are almost coincident at large $W$, where the medium differences are less dramatic.
Finally, in Fig.~\ref{pimb} we report a comparison with another model~\cite{Pimblott}. Ref.~\cite{Pimblott} uses a stochastic simulation for the partition of the different shells, and report only normalized values (hereby scaled for the present total cross section). In this case the results are slightly The major difference between the results is the slightly lower values obtained in Ref.~\cite{Pimblott} at very low energy. This region of course is still very important, thus a further improvement of detail of the energy shape, with the help of further experiments,  is desirable.
\section{conclusions}
This paper gives an opportunity to calculate the energy spectra of
secondary electrons produced by energetic ions incident on tissue,
mimiced by liquid water. In order to report these electron energy
distributions with the  best possible accuracy for each ion energy range, we consider different approaches.

The energy distributions of secondary electrons produced by low-energy ions are obtained using a general parameterization based on the Rudd Model with parameters modified for liquid water distinguished from the original parameters fitted to experiments with
water vapor. The new parameters were obtained by fitting the depths of
Bragg peaks (for different incident energies) to the experiments done
at GSI. The obtained parameterization can be used for a fast
calculation of the position of the Bragg peak for a given energy of
the projectile with a therapeutically accepted precision. An advantage
of such parameterization is its 
universality for different applications and its analyticity, which
makes all calculations fast.    

 For higher ion energies,
the dielectric response approach, previously tested for protons, and
herein applied for heavier ions, is
shown to be successful in describing the
details of secondary electron spectra profiles. 
We compare two approaches and show that they can be used for different
values of secondary electron energies.

The results of this work can be
used for calculations of realistic radial dose distributions at
different depths along the ion's trajectory. Such calculations, if
rapid enough, are important for calculating the relative biological
effectiveness and treatment planning.

Further experiments, combining more realistic projectiles (heavier
ions, since the direct ionization data are only available for protons
and helium ions~\cite{Tob80}) and more realistic media, are very much
desired.

\begin{acknowledgments} This work is partially supported by the
European Commission within the Network of Excellence project EXCELL,
the Deutsche Forschungsgemeinschaft and a FIAS fellowship. We
are grateful to D. Schardt and I. Pshenichnov  for providing original experimental and Monte Carlo data, and to A.~V. Korol and I. Mishustin for multiple fruitful discussions. J.~S. Payson's help is greatly appreciated.
\end{acknowledgments}

\end{document}